\documentclass[9pt,twocolumn,twoside]{osajnl}
\usepackage[font=small,labelfont=bf,justification=justified]{caption} % we added for justified captions 
\journal{ol} % Choose journal (ao, aop, josaa, josab, ol, pr)

\setboolean{shortarticle}{true}
\title{Using fiber-bending generated speckles for improved working distance and background-rejection in lensless microendoscopy}

\author{Noam Shekel}
\author{Ori Katz}

\affil{Department of Applied Physics, The Hebrew University of Jerusalem, Jerusalem 9190401, Israel}
\affil[*]{Corresponding author: orik@mail.huji.ac.il}

\begin{abstract}
Lensless flexible fiber-bundle based endoscopes allow imaging at depths beyond the reach of conventional microscopes with a minimal footprint. These multicore fibers provide a simple solution for widefield fluorescent imaging when the target is adjacent to the fiber facet. However, they suffer from a very limited working distance and out-of-focus background. Here, we carefully study the dynamic speckle illumination patterns generated by bending a commercial fiber-bundle, and show that they can be exploited to allow extended working distance and background rejection, using a super-resolution fluctuations imaging (SOFI) analysis of multiple frames, without the addition of any optical elements. 
\end{abstract}

\setboolean{displaycopyright}{false}

\begin{document}

\maketitle

Flexible optical micro-endoscopes are an important tool for biomedical investigations and clinical diagnostics. They enable imaging at depths where scattering prevents noninvasive microscopic investigation. An ideal micro-endoscopic probe should be flexible, allow real-time diffraction-limited imaging at various working distances from its distal end, and maintain a minimal cross sectional footprint. A robust, widely used, and commercially available type of imaging endoscope is based on a lensless fiber-bundle, which is a multicore fiber, constructed from thousands of individual cores closely packed together. Each core carries one image pixel information \cite{flusberg2005fiber,oh2013optical}. 

\begin{figure}[thpb]
\centering
\includegraphics[width=\linewidth]{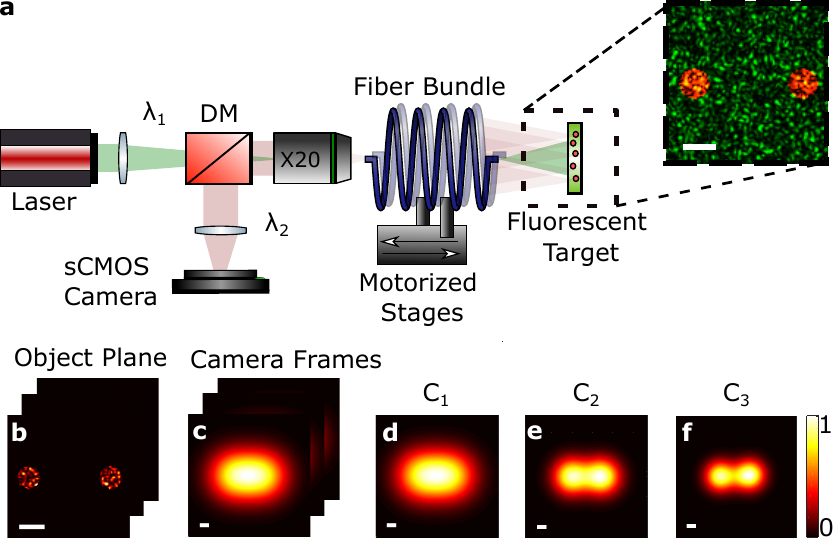}
\caption{(a) Setup. An expanded laser beam is coupled to a fiber-bundle via dichroic mirror (DM) and illuminates a fluorescent target laying at a distance from the fiber facet. Fluorescent light is collected by the bundle and imaged on an sCMOS camera. The fiber is mounted on three motorized stages, enabling controlled bending. Inset: simulated speckle pattern (green) illuminating fluorescent targets (red) at the target plane.
(b-f) Super-resolved speckle-SOFI principle (numerical results): (b) the targets at the object plane are illuminated by a series of speckle patterns. The raw images as captured by the camera (c), and their average (d) do not resolve the targets due to their distance from the bundle. High-order cumulant analysis of the image series (e-f) enable improved resolution, working-distance and background-rejection. Scale-bars (a-f): 10 $\mu m.$}
\label{fig:set_up_and_sim}
\end{figure}

The conventional setup for widefield lensless fluorescence endoscopy using a fiber-bundle is presented in Fig.\ref{fig:set_up_and_sim}(a). The setup is a simple widefield fluorescent microscope with the bundle's proximal facet at the image plane. 
Widefield imaging with a resolution limited by the cores spacing is achieved when the target object is positioned immediately adjacent to the bundle distal facet. However, the working distance is very limited, as farther away objects are blurred as each fiber core collects light from a relatively large numerical aperture ($NA$), of approximately $0.3$ in most conventional fiber bundles. The effective point spread function (PSF) width, $w(z)$, thus, grows linearly with the $NA$, and the distance from the facet, $z$ : $ w(z)\approx 2NA\cdot z $. This is a drawback in many applications, such as neuroimaging, since close contact with the fiber tip can induce tissue damage \cite{ohayon2018minimally}. An important goal is to enable wide-field high-resolution bundle-based endoscopy with larger, and potentially variable working distances.

Several solutions for lens-less bundle-based imaging with variable working distances have been demonstrated in the last couple of years. 
A simple solution for extending the working distance and depth of field (DOF) is to effectively limit the $NA$ by imaging only the fundamental mode of the bundle's cores\cite{orth2018extended}. Computational analysis of both the low- and high-order modes enable three-dimensional (3D) reconstruction and depth sectioning \cite{orth2019optical}. 
Another approach is to effectively convert the bundle to a focusing lens using wavefront-shaping. This has been demonstrated to allow 3D linear and multiphoton imaging through various bundles and multimode fibers \cite{thompson2011adaptive,bianchi2012multi,vcivzmar2012exploiting,choi2012scanner,papadopoulos2013high,andresen2013toward,rosen2015focusing,weiss2018two,kuschmierz2018self}. However, wavefront-shaping is extremely sensitive to fiber bending, as bend-induced changes in refractive index requires changing the wavefront corrections. Fiber designs that minimize the bend-sensitivity have been recently reported \cite{tsvirkun2019flexible}. 
Alternative, speckle-correlation based approaches allow diffraction-limited imaging at large distances from the fiber\cite{porat2016widefield,stasio2016calibration}. However, current works have been limited to fully developed speckles and required a minimal target distance of more than a millimeter from the fiber facet. Since scattering of biological tissue does not allow imaging at millimeters depths, using these approaches would require an additional spacer between the fiber and the sample.

In this work, we demonstrate that bend-induced dynamic refractive index variations, can be exploited for extending the working distance and reduce the inherent background in widefield lensless fiber-bundles by generating dynamic speckle-structured illumination. Speckle is observed when spatially and temporally-coherent light is transmitted through a large number of cores in conventional fiber-bundles \cite{andresen2013toward,porat2016widefield,stasio2016calibration}, in an analogue fashion to the speckle generated by light transmission through a random phase-plate \cite{goodman2007speckle}. The source of speckle is the differences in the optical paths between the bundle cores due to manufacturing tolerances, and bending-induced refractive index variations \cite{tsvirkun2019flexible}. In normal operating conditions, temporal changes in fiber orientation and temperature, dynamically change the relative phases between the cores and result in temporally-varying speckle patterns. 
While speckle may be considered a nuisance for imaging applications, several recent works have shown that dynamic speckle illumination can be used for background-rejection and super-resolution in microscopy \cite{bozinovic2008fluorescence,mudry2012structured,min2013fluorescent,kim2015superresolution}, endoscopy \cite{yin2008fluorescence} and ultrasound-mediated optical imaging \cite{mudry2012structured,chaigne2017super,murray2017super,hojman2017photoacoustic,doktofsky2020acousto}. In all of these works, the dynamic speckle illumination is generated by the addition of a controlled moving diffuser in the illumination path. 
Here, we show that the naturally dynamic speckle patterns formed by dynamic bending commercial fiber-bundles can be used to improve resolution, working distance, and background-rejection, in a similar manner to the use of diffuser-generated speckles, but without any change to the conventional optical setup (Fig.\ref{fig:set_up_and_sim}). Specifically, we exploit bending-induced dynamic speckle illumination for performing Super-resolution Optical Fluctuation Imaging (SOFI) \cite{dertinger2009fast,kim2015superresolution} in a lensless micro-endoscope. 

SOFI \cite{dertinger2009fast} is a technique that allows background rejection, and super-resolution by statistically analyzing an image-series of blinking fluorophores. To achieve super-resolution in SOFI, the $n_{th}$ order statistical cumulant, $C_{n}$, of the recorded temporal intensity fluctuations is calculated per pixel. The $n_{th}$ order cumulant image provides a $\sqrt{n}$ times resolution increase without deconvolution, and up to $n$-times resolution increase with deconvolution \cite{dertinger2009fast}. 

For example, the second cumulant given by: $C_{2}\left(r\right)=\left\langle\left(I_j\left(r\right)-\left\langle I_j\left(r\right)\right\rangle_j\right)^{2}\right\rangle_j$, where $I_j$ is a single camera frame, is simply the temporal-variance of each pixel over time, and provides a $\sqrt{2}$-fold resolution increase. Theoretically, SOFI can lead to unlimited resolution improvement, however the number of required frames, grows with the cumulant order. 

SOFI requires temporal fluctuations in the emitted fluorescence that are spatially uncorrelated at scales smaller than the PSF dimensions. Interestingly, spatially-uncorrelated fluorescence fluctuations are naturally attained by dynamic illumination of standard fluorophores with randomly fluctuating optical speckles \cite{kim2015superresolution}, forming the basis for 'speckle-SOFI' (S-SOFI). In S-SOFI, the target is illuminated by dynamic random speckle patterns, making each diffraction-limited speckle-grain spatial-position on the object to dynamically fluctuate in intensity, in a fashion analogous to a blinking emitter. Each pattern illumination is captured as a single frame in an image series (Fig.\ref{fig:set_up_and_sim}(c)). 
As in conventional SOFI, an improved resolution image is obtained in S-SOFI if the speckle diffraction-limited grain size, which is the spatial correlation of the fluctuations, is smaller than the PSF dimensions. Interestingly, this is exactly the case when considering commercial fiber bundles, as we show below. Fig.\ref{fig:set_up_and_sim}(d-f) displays sample numerical results for the gain in resolution of the first three cumulant images, by analyzing 512 frames.
In addition to the condition of the small spatial-correlation of the fluctuations, SOFI also requires the acquisition of a minimal number of uncorrelated frames to calculate the $n$-th order cumulant. Roughly, $\sim100$ realizations are required to correctly estimate $C_2$, and $\sim1000$ realizations are required for $C_3$.

To verify the feasibility of S-SOFI by fiber bending, we first quantitatively characterized the number of uncorrelated frames and the speckle grain size generated by bending of a commercial fiber bundle.
\begin{figure}[b!]
\centering
\includegraphics[width=\linewidth]{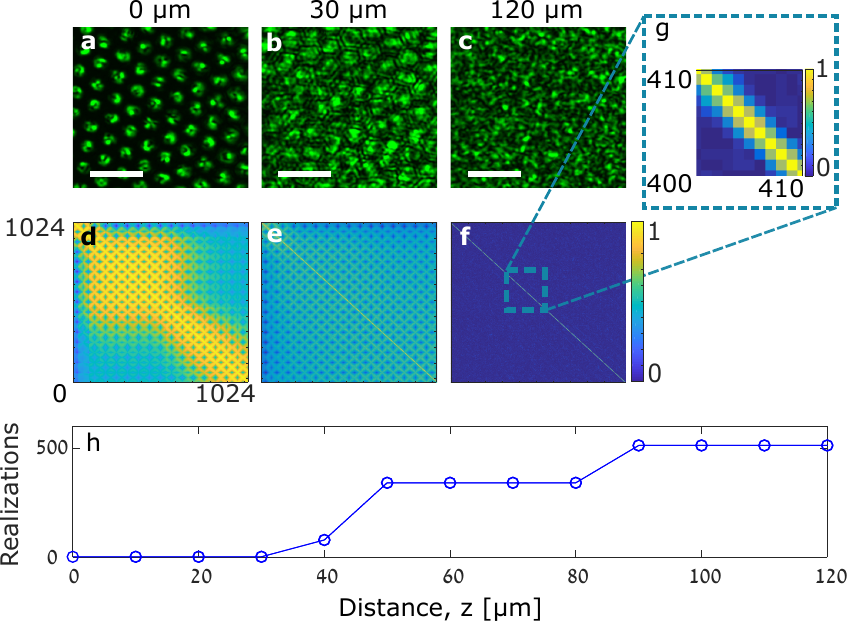}
\caption{Study of dynamic speckle patterns generated by fiber bundle bending. (a-c) raw speckle patterns measured at distances of 0,30,120 $\mu m $ respectively from the facet. (d-f) Cross-correlation matrices of $1024$ speckle patterns for each distance. (g): zoom-in on the diagonal of (f). (h) The number of uncorrelated speckle realizations (correlation < 0.25) created by fiber-bundle bending for each distance, $z$. For $z>\sim 50 \mu m $ a few hundreds independent realizations allow the use of SOFI analysis. Scale bars (a-c) 15 $\mu m $.
} 
\label{fig:correlation-vs-range}
\end{figure}

Fig.\ref{fig:set_up_and_sim}(a) displays the experimental setup used for examining the speckle patterns generated by the fiber and to demonstrate fiber-bending based SOFI. A 532 $nm$ laser beam (Standa-Q1-SH) is expanded to a diameter of $\approx 200\mu m$ and coupled to a fiber bundle (Schott P/N 1563385, pixel pitch $6.5 \mu m$) through a dichroic-mirror (DM, Omega Optical 540DRSP), and illuminates a target made from fluorescent beads (Spherotech, SPHERO FP-6056-2). The proximal side of the fiber-bundle is imaged with a 4-f system composed of an objective (Olympus PLN 20X), and a 75 $mm$ focal lens (Thorlabs LB1901-A), onto a sCMOS camera (Andor Zyla 4.2 P). The fluorescence is spectrally filtered with band-pass and long-pass filters. (Thorlabs FEL0550 + FB600-40+ Semrock BrightLine FF01-576/10-25). Dynamic speckle patterns are created in a controlled fashion by moving the fiber at three points using three motorized stages with a travel of $25mm$ (Thorlabs Z825B). To characterize the speckle patterns, an additional camera was positioned at the distal side of bundle (not shown in Fig.\ref{fig:set_up_and_sim}), and the fluorescent targets were removed.

The results of the study of the number of uncorrelated speckle patterns, calculated by cross-correlating the different captured frames, are presented in Fig.\ref{fig:correlation-vs-range}. Fig.\ref{fig:correlation-vs-range}. (a-c) are sample raw captured speckle patterns at distances of 0,30,120 $\mu m $ from the fiber facet. Fig.\ref{fig:correlation-vs-range}(d-f) are the corresponding correlation matrices, calculated for 1024 different fiber bending configurations. 
Each element in the correlation matrix represents the correlation of two speckle patterns created by fiber bending. As expected, adjacent to the bundle facet, the illumination from the different cores do not overlap, no speckles are formed, and the patterns are highly correlated Fig.\ref{fig:correlation-vs-range}. (a,d). As an estimate to the number of uncorrelated speckle patterns at each distance from the facet, we count the number of speckle patterns that have a correlation lower than $0.25$ (a threshold chosen arbitrarily).
 
Fig.\ref{fig:correlation-vs-range}.(h) summarizes the number of available independent speckle realizations as a function of the distance from the fiber facet. Our measurements show that a few hundred uncorrelated realizations are already available at distances above 50 $\mu m $.

\begin{figure}[htbp]
\centering
\includegraphics[width=\linewidth]{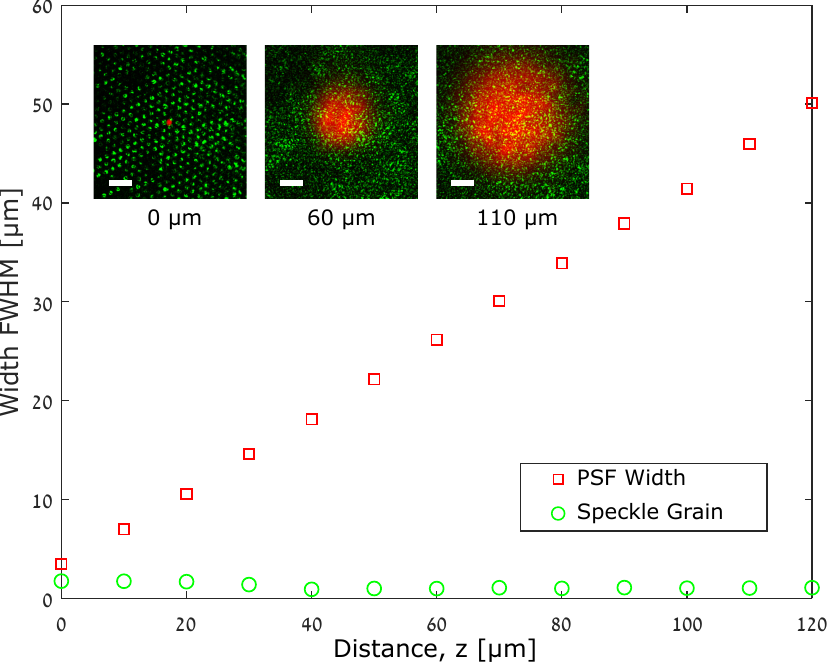}
\caption{Experimentally measured speckle grain size (green dots), and effective PSF diameter (red squares) as a function of distance form the bundle facet. Insets: images of the measured PSFs (red) super-imposed on the measured speckle patterns (green) at distances, $z$, of 0, 60 , 110 $\mu m$. For $z>\sim 10\mu m$ the speckle grain size is smaller than the PSF, allowing super-resolved imaging via SOFI. Scale bars 15 $\mu m$.
.} 
\label{fig:PSF-vs-range}
\end{figure}
The captured speckle patterns were also used to characterize the speckle grain size and compare it to the effective PSF of the bundle, at each distance. The speckle grain size was taken as the full width half max (FWHM) of the autocorrelation of each speckle intensity image. The fiber-bundle PSF was measured by placing a virtual 'point source' realized by tight focusing of a $638nm$ laser beam using a $0.65NA$ objective. At distances of $z < D_{bundle} / NA$. (all the ranges in our experiment) where $D_{bundle}$ is the diameter of the illumination on the bundle and $NA$ is the bundle's $NA$ , the speckle grain size, $\delta x$, is expected to be diffraction-limited and constant: $ \delta x =\lambda/ NA$, where $\lambda$ is the illumination wavelength \cite{porat2016widefield}, while the PSF size should grow linearly with $z$. Our experimental results of Fig.\ref{fig:PSF-vs-range} validate both expected results, and show that the speckles are considerably smaller than the PSF for all relevant ranges, as required for SOFI.

After confirming that the speckle patterns generated by fiber-bundle bending meet the requirements for S-SOFI we turned to demonstrate resolution gain and background rejection in imaging experiments, using the setup of Fig.\ref{fig:set_up_and_sim}(a). As a first test, a simple target composed of two 6.5 $\mu m$-diameter fluorescent beads was positioned at different distances from the fiber facet. For each distance, images using 100 speckle realizations were captured. In addition, the background auto-fluorescence of the fiber, without fluorescent beads, was independently measured and subtracted from the $C_1$ image, to properly compare the background-reduction of the speckle-based SOFI. Prior to the cumulant analysis the raw images, which are pixelated by the bundle cores, were Fourier interpolated.

\begin{figure}[htbp]
\centering
\includegraphics[width=\linewidth]{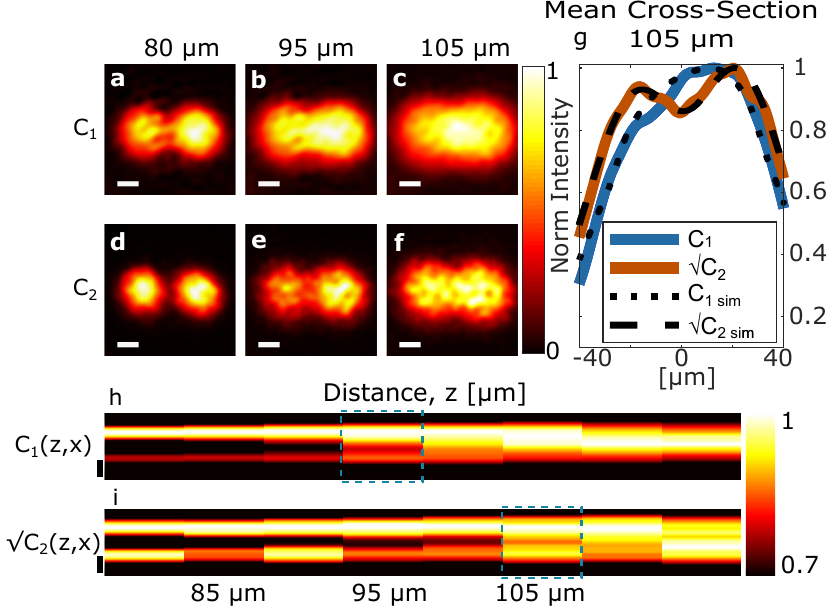}
\caption{Imaging two 6.5 $ \mu m$-diameter fluorescent beads at various distances from the bundle facet, using 100 speckle realizations for each range. (a-c) $C_1$ images at ranges 80, 95 , 105 $\mu m $ respectively. (d-f), same as (a-c) for $C_2$. (g) Mean cross-sections of $C_1$ (blue) and $\sqrt{C_2}$ (brown) taken from (c) and (f), respectively. Black squares and dashed lines are simulation results of the ideal case, showing the expected $\sqrt{2}$ increase in resolution. (h-i) Cross-sections of $C_1$ and $\sqrt{C_2}$ for ranges 80-115 $\mu m $. The $\sqrt{C_2}$ image better resolves the targets and extend the working distance by 10 $\mu m $ for this target. Scale bars (a-f, h-i) 15 $\mu m$.}
\label{fig:two_beads}
\end{figure}
Fig.\ref{fig:two_beads}(a-c) and (d-f) show the $C_1$ and $C_2$ images of the target at different distances. The $C_2$ image better resolves the beads in all distances. Fig.\ref{fig:two_beads}(g) compares the cross-section of the $C_1$ and $\sqrt{C_2}$ images when the target is at $z=105\mu m$ to a simulation of the ideal case of SOFI with a Gaussian PSF, showing an excellent agreement with the theory, and confirming the expected $\sqrt{2}$ gain in resolution. Fig.\ref{fig:two_beads}(h-i) compares the cross-sections of $C_1$ and $\sqrt{C_{2}}$ for distances of 80-115 $\mu m$. The$\sqrt{C_2}$ images better resolves the beads at all distances, extending for this target the working distance without deconvolution by 10 $\mu m$. 

\begin{figure}
\centering
\includegraphics[width=\linewidth]{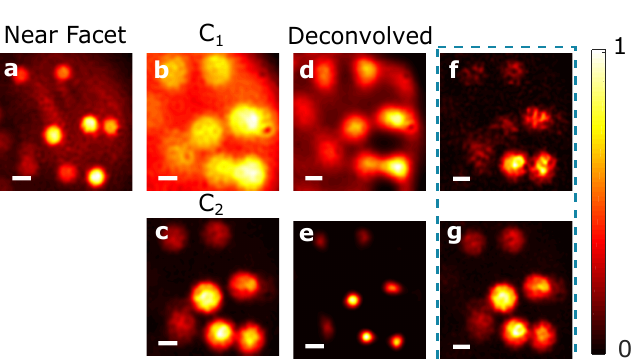}
\caption{Background reduction and deconvolution of a more complex target, placed 110 $\mu m$ from the fiber-facet. (a) Image of the target when placed adjacent to the fiber facet. (b) $C_1$ image after manual background subtraction. (c) $C_2$ image of the raw data using 100 speckle realizations, showing effective background reduction and improved resolution. (d-e) Deconvolved images of (b-c), showing improved resolution, particularly for the $C_2$ image. (f,g) $C_2$ images calculated using only 10 and 50 speckle patterns respectively, demonstrating efficient background reduction and image fidelity at shorter time-scales. Scale-bars (a-f, h-i) 30 $\mu m$. 
}
\label{fig:fig5}
\end{figure}

Fig.\ref{fig:fig5} presents the results of imaging a more complex sample with a significantly lower signal-to-background ratio, placed at a distance of $110\mu m$ from the fiber facet, using 100 speckle realizations. Fig.\ref{fig:fig5}.(b,c) present the $C_1$ and $C_2$ images, respectively, and shows that in addition to the gain in resolution, the inherent background reduction of SOFI is much better then the manual background subtraction performed for the $C_1$ image. 
Richardson–Lucy deconvolution of the $C_1$ and $C_2$ image, using a Gaussian fit of the PSF measured in Fig\ref{fig:PSF-vs-range}, are shown in Fig.\ref{fig:fig5}.(d-e). As expected, the deconvolved images show an improved resolution, and the $C_2$ deconvolved image shows a very good agreement with the target structure, measured by placing the target adjacent to the fiber facet (Fig.\ref{fig:fig5}(a)). Fig.\ref{fig:fig5}(f,g) display $C_2$ images calculated using only 10 and 50 realizations respectively, demonstrating that the background reduction, and some resolution gain, can be achieved with a rather small number of realizations, an important point for imaging speed.

To conclude, we have studied the varying speckle patterns that are generated by dynamic bending of a commercial imaging fiber-bundle. The results of Fig.\ref{fig:two_beads} and Fig.\ref{fig:fig5} show use of these speckle patterns can improve the resolution and working distance in a widefield lensless fluorescence micro-endoscope. The major difference from conventional diffuser-based speckle generation \cite{yin2008fluorescence} is that the speckle patterns are varying only from a minimal distance of approximately $50\mu m$ from the facet (Fig. \ref{fig:correlation-vs-range}(h)). Thus, the applications in lens-based fibers are not direct.

In order to best utilize the dynamic speckle illumination, a high contrast speckle should be generated. This requires the use of a sufficiently narrowband laser illumination, i.e. with a sufficiently long coherence length. Previous investigations show that for $30cm$ long fibers, the spectral bandwidth should be lower than $\sim 0.1-0.5nm $ depending on the fiber model \cite{porat2016widefield}. In our experiments, a narrowband pulsed laser was used. While its spectral bandwidth was sufficiently narrow to produce speckles with a contrast of $\sim0.7$ its high peak-power required the use of lower average power to minimize fluorescent saturation and bleaching. This limited resulted in relatively long exposure times of $\sim$1 sec to provide high signal-to-noise fluorescent images required for our study. A continuous-wave (cw) laser would enable the orders of magnitude improvement in imaging speed, required for practical applications. 

Future interesting directions may include the combination our method with the recent approach of \cite{orth2018extended}, by separately analyzing the fluctuations in the fundamental and higher-order modes, exploiting the benefits of both approaches. In addition, compressed-sensing reconstruction or blind structured-illumination-microscopy (SIM) \cite{mudry2012structured} reconstructions should also improve imaging fidelity, and reduce the number of realizations\cite{hojman2017photoacoustic}.
Coherent imaging of reflective objects with dynamic speckle illumination is also an interesting path \cite{oh2013sub}. Finally, the dynamic speckle inherently induced by bending multicore fibers, may have additional non-imaging applications, for example in fiber-based sensing. 
 
 \medskip
 \noindent\textbf{Funding.} This work is funding by the organizations: European Research Council (ERC) Horizon 2020 research and innovation program (grant no. 677909), Azrieli foundation, Israel Science Foundation (1361/18), Israeli Ministry of Science and Technology.
 
 \medskip
 \noindent\textbf{Disclosures.} The authors declare no conflicts of interest.
  
\bibliography{FB-SOFI-main}
\bibliographyfullrefs{FB-SOFI-main}

\end{document}